\documentstyle[aps,amsfonts,twocolumn,prl,floats,epsfig]{revtex}     

\newcommand{\eref}[1]{(\ref{#1})}
\newcommand{\rmd}{{\rm d}}
\newcommand{\rme}{{\rm e}}
\newcommand{\rmi}{{\rm i}}
\font\amsfnt=msam10

\begin{document}
\title{Quasilinear diffusion for the chaotic motion of a particle
       in a set of longitudinal waves%
\thanks{Presented at the 14th Marian Smoluchowski Symposium on Statistical Physics:
      Fundamentals and Applications,
      Zakopane, Poland, September 9-14, 2001}%
}
\author{D.F.~Escande and Y.~Elskens
\thanks{E-mail : x@newsup.univ-mrs.fr (x=escande, elskens).}}
\address{Equipe turbulence plasma, unit{\'e} 6633 CNRS--Universit{\'e} de Provence, \\
         case 321, Centre de Saint-J{\'e}r{\^o}me,
         F-13397 Marseille cedex 20 \\
         {\bf Acta Phys. Pol. B 33} {\rm (2002) 1073--1084} }
\date{preprint TP01.03 - {\tt arXiv.org/physics/0111206} }
\maketitle %
\begin{abstract}
The rigorous analytical calculation of the diffusion coefficient
is performed for the chaotic motion of a particle in a set of
longitudinal waves with random phases and large amplitudes ($\sim
A$). A first step proves the existence of a quasilinear diffusion
on a time scale $\sim  A^{-2/3} \ln A$. A second step uses this
property to extend the result to asymptotic times by introducing
the conditional probability distribution of position and velocity
of an orbit at a given time when they are known at a previous
time.
\newline %
PACS numbers: \newline  %
52.35.-g (Waves, oscillations, and instabilities in plasmas and
intense beams) \newline %
05.45.-a (Nonlinear dynamics and nonlinear dynamical systems) \newline %
05.60.-k (Transport processes) \newline %
52.20.-j (Elementary processes in plasmas)
\end{abstract}

\section{Introduction}
Many chaotic Hamiltonian systems encountered in physics display a
chaotic diffusion and in many cases the corresponding diffusion
coefficient is given by a so-called quasilinear estimate
\cite{VVS62,DP62,LL83,RRW79}. The proof that this estimate is
correct exists for the standard map with large control parameter
\cite{RW80}, but is lacking for other systems with a spatially
smooth force. We provide this proof for the one-dimensional
chaotic motion of a particle in a general set of waves.

This result strengthens the link between the microscopic
deterministic (chaotic) dynamics and the macroscopic stochastic
motion. Its extension to the self-consistent many-body problem is
a central problem to non-equilibrium statistical physics.

This paper is organized as follows. We first introduce our model
dynamics and stress the core of our argument. Then we recall the
traditional argument deriving the quasilinear diffusion over a
time short with respect to a characteristic time $\tau_{\rm
spread} \sim A^{-2/3}$ ($A$ being a typical value of the wave
amplitudes) and introduce the explicit form of the quasilinear
diffusion coefficient. We rederive this result within our new
approach and take advantage of a better understanding of the
particle motion to extend the validity of quasilinear diffusion to
a time scale $\sim A^{-2/3} \ln A$, which is longer than the
traditional scale $\tau_{\rm spread}$ for $A$ large. Finally, we
introduce the conditional probability distribution of position and
velocity of a chaotic orbit at a given time when they are known at
a previous time and, thanks to the non-confinement of the velocity
of the chaotic orbit, we further extend the quasilinear estimate
to asymptotic time scales.

\section{Dynamical model and assumptions}

We consider the dynamics of a particle in a set of longitudinal
waves (e.g. Langmuir waves) with random phases and large
amplitude, as defined by the Hamiltonian
\begin{equation}
  H(p,q,t)
  =
  {p^2 \over 2}
  + \sum_{m=1}^{M} A_m \cos(k_m q - \omega_m t + \varphi_{m}),
  \label{Hkomega}
\end{equation}
where the $\varphi_{m}$'s are random variables, and the
$(A_m,k_m,\omega_m)$'s are prescribed triplets of positive
parameters. Such a dynamical system has already been studied in
the literature, and for large $A_m$'s the diffusion coefficient
has been found numerically to take on the quasilinear value
\cite{RRW79,CEV90,IXW93,BE97,R98} defined below \cite{NoteA}. The
average over $M \gg  A^{2/3} \ln A \gg 1$ random phases is central
to our proof, in agreement with the occurrence of uncontrolled
phases in many experiments and with the fact that the transport in
\eref{Hkomega} is much less diffusion-like if one averages only
over initial conditions $(p_0, q_0)$ \cite{BE98}. The large $A$
limit (dynamically speaking, the limit of strong resonance overlap
parameter) corresponds to the limit of continuous spectrum often
encountered in physics.

In agreement with most of the literature on quasilinear transport,
the analysis is performed here in terms of quadratic means, and
not in terms of the probability distribution functions, but we
indicate at the end of this paper how our technique could be used
to prove the gaussianity of such functions.

The equations of motion are %
\begin{eqnarray}
  \dot q &=& p  ,
  \\
  \dot p &=& \sum_{m=1}^{M} A_m k_m
             \sin(k_mq-\omega_m t+\varphi_{m}).
\end{eqnarray}
We first consider the time to be
short enough for the orbit to stay close to the unperturbed orbit
$q^{(0)}(t)=q_0+p_0 t$, and let $ \Delta q(t)= q(t) - q^{(0)}(t)$,
$\Delta p(t) = p(t) - p_0$. We compute their statistical
properties when averaging over all $\varphi_{m}$'s.

For completeness, we first evaluate $\Delta p(t)$ by the
traditional technique \cite{VVS62,DP62} using first order
perturbation in the amplitudes:
\begin{eqnarray}
  \Delta p(t)
  &=&
  \sum_{m=1}^{M} {(A_m k_m / \Omega_m)}
  \cr
  &&
    [\cos(k_m q_0 +\varphi_m) - \cos(\Omega_m t +k_m q_0 +\varphi_m)],
\end{eqnarray}
with $\Omega_m = k_m p_0 -\omega_m$; if $\Omega_m = 0$
for some $m$, the corresponding term in the sum is the limit as
$\Omega_m \to 0$. At this order, $\langle\Delta p(t)\rangle=0$ and
$\langle\Delta p^2(t)\rangle= \sum_{m=1}^{M} ({A_m k_m /
\Omega_m})^2[1-\cos(\Omega_m t)]$.

Let $v_m = \omega_m / k_m$. We assume that $\Delta \Omega_m =
\Omega_{m+1} - \Omega_m$ and $\Delta v_m = v_{m+1} - v_m$ have a
sign independent of $m$, which is natural for Langmuir waves and
for the dynamics of Ref. \cite{BE97}. Let
\begin{equation}
  D_m
  \equiv
  {\pi A_m^2 k_m \over 2 |\Delta v_m|}
  =
  \lim_{p_0 \to v_m}
  {\pi (A_m k_m)^2 \over 2 |\Delta \Omega_m|} .
\label{Diffm}
\end{equation}
$D_m$ may fluctuate with $m$, but we assume (for simplicity only)
that for some $L \geq 0$, $\sum_{j=-L}^{L} D_{m+j}|\Delta
v_{m+j}|/|v_{m+L+1} - v_{m-L}|$ is a constant $D_{\rm QL}$, called
the quasilinear diffusion coefficient. Let $\Delta \Omega_{LM} =
\max |\Omega_{m+L+1} - \Omega_{m-L}|$, $\tau_{\rm discr} = \Delta
\Omega_{LM} ^{-1}$ and $\tau_{\rm c} =
(\Omega_{\max}-\Omega_{\min})^{-1}$ ; $\tau_{\rm discr}$ and
$\tau_{\rm c}$ are respectively the discretization time and the
correlation time of the wave spectrum as seen by the particle.

\section{Non-chaotic initial quasilinear transport}

Assuming $\tau_{\rm c} \ll t \ll \tau_{\rm discr}$, one obtains
$\langle \Delta p(t)^2 \rangle = (2 D_{\rm QL} / \pi)
\int_{-\infty}^{\infty} \Omega^{-2} [1-\cos(\Omega t)] \rmd \Omega
= 2 D_{\rm QL} t$, where the discrete sum has been turned into an
integral. As a result, the diffusion coefficient takes on the
quasilinear value $D_{\rm QL}$. A similar calculation for $q$
yields $\langle \Delta q(t) \rangle = 0$ and $\langle \Delta
q(t)^2 \rangle = 2 D_{\rm QL} t^3 / 3$. For $t \ll \tau_{\rm c}$,
$\Delta p$ grows linearly with time, and $\langle \Delta p^2
\rangle$ grows quadratically, as all modes act with a constant
force on the orbit. For $\tau_{\rm c} \ll t \ll \tau_{\rm discr}$,
the range of $m$ contributing to the diffusion (modes acting with
a nearly constant force) narrows like $1/t$. The range of $t$ is
further restricted by the condition for the orbit to remain close
to the unperturbed one. This is traditionally obtained by
requiring $\langle k_{\max}^2 \Delta q^2 (t) \rangle  \ll 4\pi^2$,
namely $t \ll \tau_{\rm spread}$ with
\begin{equation}
  \tau_{\rm spread}
  =
  \Bigl( 6 \pi^{2} k_{\max}^{-2} D_{\rm QL}^{-1} \Bigr)^{1/3}
  =
  4 \gamma_{\rm D}^{-1}
\label{tausp'}
\end{equation}
where we introduce the resonance broadening frequency
$\gamma_{{\rm D} n} \equiv (k_n^2 D_{\rm QL})^{1/3}$ and take
$\gamma_{{\rm D}} \equiv \max_n \gamma_{{\rm D}n}$.

In our approach, we evaluate $\Delta p(t)$ as in Ref. \cite{BE97}
by integrating formally the equation of motion for $p$. This
yields $\langle \Delta p(t) \rangle = 0$ over the range $0 \leq t
\ll \tau_{\rm QL}$ defined below, and $\langle \Delta p^2(t)
\rangle = \Delta_0 + \Delta_+ + \Delta_-$, with
\begin{eqnarray}
  \Delta_j
  &=&
  - \eta_j  \int_0^t \int_0^t
  \sum_{m_1=1}^{M} \sum_{m_2=1}^{M}
  {A_{m_1} k_{m_1} A_{m_2} k_{m_2} \over 2}
  \cr
  &&
  \langle \cos[\Phi_{m_1}(t_1) + \eta_j \Phi_{m_2}(t_2)] \rangle
  \rmd t_1 \rmd t_2
  \label{e14b}
\end{eqnarray}
where $\Phi_{m}(t) = k_m \Delta q(t) + \Omega_m t + \varphi_m$,
with $\eta_{\pm} = \pm 1$ and $\eta_0 = -1$, and under condition
$m_1 \neq m_2$ for $j=-$, and condition $m_1 = m_2$ for $j=0$. Let
$t_- = t_1-t_2$ and $t_+ = (t_1+t_2)/2$. For $t_- \ll \tau_{\rm
spread}$, $\langle \exp [\rmi k_m \bigl(\Delta q(t_+ + t_-/2)-
\Delta q(t_+ - t_-/2)\bigr)] \rangle$ may be considered as equal
to 1. Therefore the support in $t_-$ of the integrand in
$\Delta_0$ is of the order of $\tau_{\rm c}$. We assume $\tau_{\rm
c} \ll \tau_{\rm spread}$. Hence the integration domain in $t_-$
may be restricted to $|t_-| \leq \nu \tau_{\rm c}$ where $\nu$ is
a few units. In the limit where $\nu \tau_{\rm c} \ll  t \ll
\tau_{\rm discr}$, we obtain %
$\Delta_0
  = \sum_{m=1}^{M} \int_0^t
    ({2 D_m / \pi}) \int_0^{\nu \tau_{\rm c}}
    \langle \cos[\Omega_m t_-] \rangle
    \Delta \Omega_m \rmd t_- \rmd t_+
  = 2 D_{\rm QL} \sum_{m=1}^{M}
    (\pi \Omega_m)^{-1}
    \langle \sin[\Omega_m \nu \tau_{\rm c}] \rangle
    \Delta \Omega_m t
  = 2 D_{\rm QL} t$,
with the discrete sum over $m$ approximated by an integral.

For $t \ll \tau_{\rm spread}$ we approximate $q(t)$ by its
unperturbed value $q^{(0)}(t)$. As this orbit does not depend on
the phases, the averaged cosines in \eref{e14b} are zero for $j =
\pm$, and so are the $\Delta_\pm$'s. Then our second approach
shows again that the diffusion coefficient takes on the
quasilinear value. $\langle \Delta q^2 (t) \rangle$ too may be
computed by integrating the equation of motion \cite{NoteQ}. This
involves calculating $\langle \Delta p(t_1) \Delta p(t_2)
\rangle$, in the same way as $\langle \Delta p^2 (t) \rangle$, and
one recovers the traditional estimate for $\langle \Delta q^2 (t)
\rangle$. This provides a way for introducing the condition $t \ll
\tau_{\rm spread}$ without resorting to the traditional
perturbative approach, and shows that the usual quasilinear
diffusion coefficient may be recovered independently by our second
approach.

\section{Chaotic trajectory spreading}

In fact our second approach is much more powerful. As was pointed
out in Ref. \cite{BE97}, $\Delta_\pm$ vanishes provided that the
dependence of $\Delta q$ over any $N_\varphi = 2$ phases with all
other phases fixed is weak, a condition far less stringent than
the previous condition $N_\varphi = M$ which led to $t \ll
\tau_{\rm spread}$. Reference \cite{BE97} estimated the upper
bound in time of the initial quasilinear diffusion through
numerical calculations for moderate values of the waves amplitude.
Here we derive such a bound analytically for large enough
amplitudes.

We measure these amplitudes by the parameter $E_n = [2 D_{\rm QL}
k_n |\Delta v_n|/ \pi]^{1/2}$ which corresponds to the typical
electric field of a wave. A related dimensionless quantity
characterizes our scaling, namely the Chirikov resonance overlap
parameter
\begin{equation}
  s(v_n)
  =
  2 [A_n^{1/2} + A_{n+1}^{1/2}]/|\Delta v_n|
  \label{soverlap}
\end{equation}
or equivalently the ratio  %
\begin{equation}
  {\cal B} (v_n)
  \equiv k_n |\Delta v_n| / \gamma_{{\rm D}n}%
  \simeq 5 s^{-4/3}   %
\label{defBrosd}
\end{equation}
of the frequency mismatch between neighbouring waves (in the frame
of either wave) to their resonance broadening frequency. As these
quantities depend on $n$, they characterize the dynamics locally.
In the following, we are interested in the dense spectrum, or
strong overlap, or large amplitude limit. To ensure a genuine
scaling, we consider families of dynamics \eref{Hkomega} where
$E_n = E a_n$ and the reference amplitudes $a_n$ are constant
while $E \to \infty$, or ${\cal B}(v_n) = {\cal B}b_n$ and the
coefficients $b_n$ are constant while ${\cal B} \to 0$.

Apart from the small dimensionless parameter ${\cal B}$, we also
introduce the Kubo number ${\cal K}_{\rm c} \equiv \tau_{\rm c} /
\tau_{\rm spread}$. The wide velocity spectrum of the waves
ensures that ${\cal K}_{\rm c} \ll 1$.

The limit of interest is the {\bf joint limit} ${\cal K}_{\rm c}
\to 0$ and ${\cal B} \to 0$ (or ${\cal K}_{\rm c} \to 0$ and $s
\to \infty$).

\subsection{Spreading due to a single random phase}

In order to avoid too heavy formulas, we give the explicit
derivation for the spreading due to one phase, and extend the
result to two phases afterwards. To estimate this spreading we
study how the orbit which is at $(q_0,p_0)$ at $t=0$ is modified
when phase $\varphi_{n}$ changes from 0 to a finite value. Let
$(q_{\not n}(t), p_{\not n}(t))$ be the orbit for $\varphi_{n}=0$,
let $\delta q_n(t) = q(t) - q_{\not n}(t)$ and $\delta p_n(t) =
\delta \dot q_n(t) = p(t) - p_{\not n}(t)$. We assume $t$ to be
small enough so that $k_{\max} |\delta q_n(t)| \ll \pi$. As
$\delta q_n(t)$ is small, we may linearize the motion
\begin{equation}
  \delta \dot p_n(t)
  \simeq
  F(t) \delta q_n(t)
  + A_n k_n (\sin \Psi_n(t) - \sin \Psi_{n0}(t))
  \label{e3}
\end{equation}
where $F(t) = \sum_{m=1}^{M} k_m^2 A_m\cos \Psi_m(t)$, with
$\Psi_m = k_m q_{\not n}(t) - \omega_m t + \varphi_m$ and
$\Psi_{n0} = k_n q_{\not n}(t) - \omega_n t$. Then \eref{e3} and
initial conditions $(\delta q_n(0), \delta p_n(0)) = (0,0)$ imply %
  \begin{equation}
    \delta q_n(t)
    =
    \int_0^t (t - t'') F(t'') \delta q_n(t'') d t''
     + \delta q_{n0}(t)
  \label{dqnt}
  \end{equation}
where %
$\delta q_{n0}(t)
  = A_n k_n \int_0^t \int_0^{t'}
   \bigl( \sin \Psi_{n}(t'') - \sin \Psi_{n0}(t'') \bigr)
   \rmd t'' \rmd t'$.
In the short-time limit, the dominant term in expression
(\ref{dqnt}) for $\delta q_n$ will be $\delta q_{n0}$, but over
longer times the first term may self-amplify and overtake the
second one.

We only estimate $\langle \delta q_n(t)^2 \rangle$, but $\langle
\delta q_n(t) \rangle$ can be computed by the same technique and
turns out to be negligible over the time interval of interest. In
a first stage, consider the contribution of $\delta q_{n0}$
to the variance, %
$C_{0}(t)
  \simeq
  \langle \delta q_{n0}(t)^2 \rangle
  = (k_n^2 A_n^2 / 2)
  \int_0^{t} \int_0^{t'_1} \int_0^{t} \int_0^{t'_2}
    \bigl\langle \cos \bigl( \Psi_{n}(t''_1) - \Psi_{n}(t''_2) \bigr)
    \bigr\rangle
    \rmd t''_2 \rmd t'_2 \rmd t''_1 \rmd t'_1$.
To estimate this expression, note that $\Psi_n(t''_1) -
\Psi_n(t''_2) - \Omega_n (t''_1 - t''_2) = k_n (q_{\not n}(t''_1)
- q_{\not n}(t''_2)) - k_n p_0 (t''_1 - t''_2)$, and, for the
range of time of interest, $q_{\not n}(t'') - q^{(0)}(t'')$ is
essentially the sum of $M-1$ terms in which a random phase
$\varphi_m$ ($m \neq n$) is added to a term which has a weak
dependence on $\varphi_m$. Therefore, this sum is almost gaussian,
and for $M \gg 1$ we may approximate $\dot q_{\not n}(t'')$ by a
brownian motion. Furthermore, as $M \gg 1$, we approximate
$q_{\not n}(t'')$ by $ q(t'')$ in the averages. Using the
distribution of $\Delta q (t_2) - \Delta q (t_1)$, we find
\cite{NoteC}
the estimate %
\begin{equation}
  C_{0}(t)
  \leq
  C_{0{\rm M}}(t)
  \equiv
  0.28  k_n |\Delta v_n|   \gamma_{{\rm D}n}^2 t^3
  =
  0.28 {\cal B} \ (\gamma_{{\rm D}n} t)^3 .
\end{equation}

For the second stage, we take into account the first term in the
right hand side of \eref{e3}. As $\delta q_n$ is small, we may
treat $F(t)$ as a gaussian process with moments $\langle F(t)
\rangle = 0$ and $\langle F(t_1) F(t_2) \rangle = 2 \gamma_{{\rm
D}n}^3  \delta (t_1 - t_2)$ where $\delta (t)$ is the Dirac
distribution. Indeed $q_{\not n}(t)$ has a weak dependence on any
phase $\varphi_m$, which makes $\langle F(t_1) F(t_2) \rangle$ a
Bragg-like function with the small width $\tau_c$ in $t_1 - t_2$.
Higher moments of $F$ are assumed to factorize, i.e. $F$ is
treated as a white noise, which is consistent with approximating
$\dot q_{\not n}(t)$ by a brownian motion.

We estimate the spreading of $\delta q_n(t)$ by computing %
\begin{eqnarray}
  C(t)
  & \equiv &
  \langle \delta q_n(t)^2 \rangle
  \cr
  & \simeq &
  \int_0^t \int_0^{t'_1} \int_0^t \int_0^{t'_2}
      \langle F(t''_1) F(t''_2) \rangle
      \langle \delta q_n(t''_1) \delta q_n(t''_2) \rangle
  \cr
  && \hskip1cm
      \rmd t''_2 \rmd t'_2 \rmd t''_1 \rmd t'_1
  +
  C_{0}(t)
  \cr
  &=&
  (E^2 / 2)
  \int_0^{t_1} \int_0^{t_2} \int_0^{\min(t'_1, t'_2)}
    C(t'') \rmd t'' \rmd t'_2 \rmd t'_1
  \cr
  &&\hskip1cm
  +
  C_{0}(t).
\label{decompC}
\end{eqnarray}
It follows from \eref{e3} and our assumptions on $F$ that %
$C(t) = C_0(t) + L C(t)$ with \par \noindent%
$L f(t)  =
  (E^2 / 2)
  \int_0^{t} \int_0^{t} \int_0^{\min(t'_1, t'_2)}
   f(t'') \rmd t'' \rmd t'_2 \rmd t'_1$. \par\noindent %
As $(1-L)^{-1}$ preserves positivity \cite{NoteL}, %
$C = (1-L)^{-1} C_0 \leq (1-L)^{-1} C_{0{\rm M}} \equiv C_{\rm
M}$. Applying the Laplace transform to both sides of equation
$C_{\rm M} = C_{0{\rm M}} + L C_{\rm M}$, we compute $C_{\rm M}$
and find
\begin{equation}
  C(t)
  \leq
  C_{\rm M}(t)
  =
  0.14 {\cal B} k_n^{-2}
  \bigl( \rme^{t'} - 1 + 2 g(t')
         \bigr)
\label{e507}
\end{equation}
with $t' \equiv 4^{1/3} \gamma_{{\rm D}n} t$ and $g(t') = \rme^{-
t' /2 } \cos(t' \sqrt{3}/2) - 1$. This estimate for the variance
of $\delta q_n(t)$ starts from zero at $t=0$ and diverges
exponentially for $t \to \infty$. Its exponentiation time scale
$\tau_{\rm Liap} \sim \gamma_{{\rm D}n}^{-1} \sim \tau_{\rm
spread}$ is the reciprocal of the Liapunov characteristic
instability rate (this is reminiscent of Ref. \cite{RRW79}).
However, as the coefficient in front of the exponential goes to
zero as $E \to \infty$, the time needed by our upper estimate on
$k_n^2 C(t)$ to reach unity is of the order of
\begin{equation}
  \tau_{\rm QL}
  =
  \gamma_{{\rm D}}^{-1}  | \ln {\cal B} |
  \label{timeQL}
\end{equation}
Though this time goes to zero as $E \to \infty$, it is ${\rm
O}(\ln {\cal B}^{-1})$ times larger than the time $\tau_{\rm
spread}$ over which the initial quasilinear approximation is
traditionally justified.

\subsection{Spreading due to two random phases}

The result of this discussion is that ``$q(t)$ depends little on
any given phase over a time $\tau_{\rm QL}$''. For $M \gg 1$, the
argument is easily strengthened into ``$q(t)$ depends little on
any two given phases over a time $\tau_{\rm QL}$''. To this end
$(q_{\not m_1, \not m_2}(t),p_{\not m_1, \not m_2}(t))$ and
$(\delta q(t), \delta p(t))$  are defined starting from
$\varphi_{m_1} = \varphi_{m_2} = 0$, and a third term similar to
the second one adds in the right hand side of \eref{e3}. The first
stage of our iteration procedure now estimates the contribution of
both phases $\varphi_{m_1}$ and $\varphi_{m_2}$ by a term again of
the order of ${\cal B}^{-2} \gamma_{{\rm D}}^3 t^3$, while the
second stage does not change.

As a result, for $t \ll \tau_{\rm QL}$, the non-quasilinear terms
$\Delta_\pm$ are negligible since $q$ has a small dependence on
any given pair of phases in this time range. Furthermore these
terms may be estimated by expliciting in the argument of the
cosine of \eref{e14b} the main dependence over $\varphi_{m_1}$ and
$\varphi_{m_2}$ through estimates $\delta \Phi_{m_1}$ and $\delta
\Phi_{m_2}$ of the type $k_m \delta q_{n0}$ for both phases, and
by expanding to second order in these $\delta \Phi$'s. Such
estimates hold for $t \ll  \beta \tau_{\rm QL}$ with $0< \beta <
1$ for $E$ large enough, and yield $\Delta_+ \sim E^4 \tau_{\rm
c}^2 t^2$ and $\Delta_- \sim E^4 t^5$ which are negligible with
respect to $\Delta_0$ in the time interval of interest, and do not
grow with $M$ although there are $2M^2-M$ ``off-diagonal'' terms.

\section{Quasilinear transport over large times}

Finally, we show that the quasilinear estimate holds for
asymptotic times. Let $p_{\min} = \min(v_m)$ and $p_{\max} =
\max(v_m)$. We assume that in the velocity domain
$[p_{\min},p_{\max}]$ the dynamics is chaotic enough for a typical
orbit to be unconfined in $p$ within this domain, but that the
time of interest is also smaller than the time for the orbit to
reach the boundaries of the chaotic domain. Therefore we set
the condition %
  $\min[(p_0-p_{\min})^2,(p_0-p_{\max})^2]
   \gg
   D_{\rm QL} \tau_{\rm QL}
   \sim
   k_n^{-2} \gamma_{{\rm D}n}^2 \ln ({\cal B}^{-1})$ %
to compute now the diffusion coefficient due to the chaotic motion
when $M$ and $E$ are large. We define $\delta q (\tau|p,q,t) =
q(t+\tau) - q - p \tau$, where $q(t')$ is the position at time
$t'$ of an orbit which is at $(p,q)$ at time $t$~: $\delta q
(\tau|p,q,t)$ tells the departure of this orbit from the free
motion during the time interval $\tau$.

Integrating formally the equation of motion for $p$ yields
\begin{equation}
  \langle \Delta p^2(t) \rangle
  =
  - \sum_{m,n=1}^{M} \sum_{\epsilon=\pm 1}
    \epsilon {A_m k_m A_n k_n \over 2} \int_0^t \int_0^t
    \langle \cos \Phi \rangle
    \rmd t'\rmd t''
\label{Dp2k}
\end{equation}
where $\Phi = (k_m + \epsilon k_n) q(t'') + k_m \delta q[t'-t''
|p(t''),q(t''),t''] + k_m p(t'')(t'-t'') -\omega_m t' -\epsilon
\omega_n t'' +\varphi_m + \epsilon \varphi_n$. We introduce the
probability distribution $P(\delta p,t|p_0)$ of $\delta p =
p(t)-p_0$ for an orbit started at $p=p_0$ at $t_0=0$; it is
independent of $q_0$.

$\langle \cos[k_m \delta q
\bigl(t'-t''|p(t''),q(t''),t''\bigr)]\rangle$ is independent of
$q(t'')$, and its contribution for diagonal ($m=n$, $\epsilon=-1$)
terms to \eref{Dp2k} is
\begin{eqnarray}
  &&
  B
  \cr
  & \equiv &
  \lim_{t \to \infty} \sum_{m=1}^{M}{(A_m k_m)^2\over 4t}
     \int_0^t \int_0^t \int
     P(\delta p,t''|p_0)
  \cr
  & &
    \hskip5mm
     \langle \cos[k_m \delta q \bigl(t'-t''|p_0,q(t''),t''\bigr)] \rangle_*
  \cr
  & &
    \hskip5mm
     \cos[k_m (p_0 + \delta p)(t'-t'')-\omega_m (t' - t'')]
  \cr
  & &
    \hskip5mm
     \rmd \delta p \rmd t'\rmd t''
  \cr
  & = &
  \lim_{t \to \infty}
     \sum_{m=1}^{M} {(A_m k_m)^2 \over 4 t} \Re
     \int_0^t \int_{-t''}^{t-t''}
      \tilde{P}(k_m \tau,t''|p_0)
  \cr
  & & \hskip5mm
       \exp[\rmi \Omega_m \tau]
       \langle \exp [\rmi k_m \delta q (\tau|p_0,q(t''),t'')] \rangle_*
     \rmd \tau \rmd t'' ,
\label{cosav}
\end{eqnarray}
where the starred average means the average done with the
constraint $p(t'')=p_0 + \delta p$, and where the Fourier
transform
\begin{equation}
  \tilde{P}(\alpha,t''|p_0)
  =
  \int_{-\infty}^{\infty}
    P(\delta p,t''|p_0) \exp (\rmi \alpha \delta p) \rmd \delta p
\label{FourierP}
\end{equation}
was used. As $\delta q$ is computed with the knowledge of $p$ at
time $t''$ which sets only one condition on a set of many phases,
an average with the constraint $p(t'')=p_0 + \delta p$ may be
computed by using the initial quasilinear estimate at time
$|t'-t''| \leq \tau_{\rm QL}$. Hence the function $\langle
\exp[\rmi k_m \delta q (t'-t''|p,q(t''),t'')]\rangle_*$ is
correctly computed by the previous quasilinear estimate over its
whole support in $t'-t''$ as $\tau_{\rm QL} \gg \tau_{\rm
spread}$. This estimate is independent of $p$, and we could set
$p=p_0$ in the average cosine. Up to $t=\tau_{\rm QL}$, the width
of $P$ is growing, since we proved $\langle \Delta p^2(t) \rangle$
grows linearly over this time interval. Later on this width cannot
decrease because of the locality of chaotic motion
\cite{BE97,BE98b}. We assume $t \gg \tau_{\rm spread}$. Then the
width $w$ of $\tilde{P}$ is narrow enough for the spread of
$\delta q$ to be negligible over a time $\tau \sim w /k_m$.
Therefore $\langle \exp[\rmi k_m \delta q (\tau|p_0,q(t''),t'')]
\rangle_* \simeq 1$ in the part of the integration domain over
$\tau$ where $\tilde{P}$
takes appreciable values in \eref{cosav}, and %
$B =
  \lim_{t \to \infty} \sum_{m=1}^{M}
  {\pi A_m^2 k_m \over 2t} \int_0^t
    P(v_m-p_0,t''|p_0)
    \rmd t''
 = \int_0^t \sum_{m=1}^{M}{D_m \Delta v_m \over 2t}
   P(v_m-p_0,t''|p_0) \rmd t''$, %
where the inverse Fourier transform was provided by the integral
over $\tau$.

Now, if $t$ is large enough for $P$ to be almost constant over the
range $[v_{m-L}, v_{m+L}]$ for all $m$'s, we approximate
$\sum_{j=-L}^L D_{m+j} |\Delta v_{m+j}| / |v_{m+L+1} - v_{m-L}|
\simeq  D_{\rm QL}$
and substitute the sum over $v_m$ by an integral : %
$B =
  2 \int_0^t \int D_{\rm QL} P(p-p_0,t''|p_0) \rmd p \rmd t''
  = 2 D_{\rm QL} t$.

The general term of \eref{Dp2k} can be estimated by a similar
calculation. A sequence of two Fourier transforms is again
recovered. After the first one, averages of the kind $\langle \exp
\rmi[k_m \delta q (\tau|p_0,q(t''),t'') + \varphi_m + \epsilon
\varphi_n] \rangle_*$ are found. They vanish as the constraint
$p(t'') = p_0 + \delta p$ leaves almost free the average on any
two phases, and since $\delta q$ is negligible for $\tau$ small.
Therefore only $B$ contributes to $\langle \Delta p^2(t) \rangle$
which thus grows in a quasilinear way. This ends our proof of the
quasilinear estimate for asymptotic times.

Note that the conditional probability $P$ permits to use the
knowledge of initial quasilinear diffusion for proving it over
asymptotic times only because we proved before that $\tau_{\rm QL}
\gg \tau_{\rm spread}$. In contrast with the initial non-chaotic
quasilinear regime, the number of modes acting on the particle
increases with $t$. This agrees with the fact that the orbit
visits an increasing number of resonances when time increases.

\section{Conclusion}

Thus we prove the quasilinear character of the diffusion for the
motion of a particle in a spectrum of large amplitude longitudinal
waves. Our technique can be adapted to systems with a slow
dependence of the quasilinear diffusion coefficient on $p$. As
many Hamiltonian systems may be locally reduced to case
\eref{Hkomega} \cite{E85}, this further extends its range of
applicability and shows that the universality class of quasilinear
diffusion is broad. It also provides insight for the case where
particles and waves are self-consistently coupled \cite{DC97}.

Higher order moments of $\Delta p$ could be computed using a
similar technique. Indeed, preliminary calculations indicate that
the use of conditional probabilities should enable one to retain
after Fourier transforms the same terms for the moment of order
$\kappa$ as in the case where $q(t)$ is weakly dependent on any
phase provided that $\kappa \ll {\cal B}^{-1}$, which yields a
gaussian estimate. Proving the Gaussianity of $f$ would also lead
to a Fokker-Planck-Smoluchowski evolution equation for $f$.

The value of ${\cal B}$ (which depends only on local aspects of
the spectrum : $A$, $k$, $\delta v$) determines the time scale
over which the quasilinear approximation holds. Given ${\cal B}
\ll 1$, this time scale is $t \gg \tau_{\rm QL}$. On the other
hand, we require that the motion remains away from the boundaries
$p_{\min}$ and $p_{\max}$ of the wave spectrum. Given the scaling
$\langle \Delta p^2 \rangle  \sim  2 D t$, the boundary is reached
for $t_{\rm bound} \sim D^{-1} M^2 \Delta v^2 \sim M^2 {\cal B}
\tau_{\rm QL}$. As $M$ is independent of ${\cal B}$, one may let
$M \to \infty$ to ensure $t_{\rm bound}$ to be as large as
desirable.

\vskip10pt

Comments by D. B\'enisti and A. Henriet on this work are
gratefully acknowledged. YE thanks the organizers of the M.
Smoluchowski symposium for discussions.





\begin{thebibliography}{99}
\bibitem{NoteA} {$D_{\rm QL} = \pi A^2/2$ in the case of
  Ref. \cite{BE97} (where $M = 2M' + 1$, $A_m = A$, $k_m = 1$
  and $\omega_m = m - M' - 1$ for all $m$'s) and for the standard map,
  which is a special case of Ref. \cite{BE97} in the limit
  $M \to \infty$ with all phases $\varphi_m=0$. }
\bibitem{NoteQ} {The process $\Delta q(t)$ is found to be gaussian,
  and its moment generating function reads
  $\langle \rme^{\rmi u (\Delta q(t_2) - \Delta q(t_1)) } \rangle
   =
   \rme^{- u^2 D_{\rm QL}[|t_2 - t_1|^3/3 + |t_2 - t_1|^2
   \min (t_1,t_2) ]}$.}
\bibitem{NoteL} {
  Indeed, $L$ preserves positivity and is a contraction operator
  for functions on $[0, \infty[$ with the norm
  $\Vert f \Vert_\lambda =
   \lambda\int_0^\infty \rme^{-\lambda t} |f(t)| \rmd t$,
  for any $\lambda > 4^{1/3} \gamma_{\rm D}$.}
\end{thebibliography}
\end{document}